\documentclass[12pt]{article}
\usepackage{graphics}
\input epsf

\newcommand{\be}{\begin{eqnarray}}
\newcommand{\ee}{\end{eqnarray}}

\begin{document}
\title{
\vskip 1cm
Problems in a weakless universe}
\author{L.\ Clavelli\footnote{lclavell@bama.ua.edu}\quad and R.\ E.\ White III
\footnote{rwhite@bama.ua.edu}\\
Department of Physics and Astronomy\\
University of Alabama\\
Tuscaloosa AL 35487\\ }
\maketitle
\begin{abstract}
    The fact that life has evolved in our universe constrains the laws of physics.
The anthropic principle proposes that these constraints are sometimes very tight
and can be used to explain in a sense the corresponding laws.        
Recently a ``disproof" of the anthropic principle has been proposed \cite{Harnik} 
in the form of a universe without weak interactions, but with other parameters 
suitably
tuned to nevertheless allow life to develop.  
If a universe with such different physics from ours can generate life,
the anthropic principle is undermined.
We point out, however, that on closer
examination the proposed ``weakless" universe strongly inhibits the development
of life in several different ways.  One of the most critical barriers is that
a weakless universe is unlikely to produce enough oxygen to support life.
Since oxygen is an essential element in both
water, the universal solvent needed for life, and in each of the four bases
forming the DNA code for known living beings, we strongly question the hypothesis
that a universe without weak interactions could generate life. 
\end{abstract}
PACS number:12.60.-i 

\section{\bf Introduction}
\setcounter{equation}{0}
    In the last few years the anthropic principle (AP) has caused some turmoil in
the physics and astronomy communities.  It cannot be disputed that the fact
that we are here making observations on nature implies that the laws of
physics are such as to have made this possible (observer bias).  The problem
perhaps comes from the prospect that the AP can be used as an explanation
for the tightly constrained aspects of physical law.  Certainly the
principle does not provide an explanation in the traditional sense nor
suggest that explanations of the traditional type do not generally exist. 
Nevertheless there is widespread discomfort in the science community
typified perhaps by the remarks of Burton Richter at the June 2006 conference
on Supersymmetry \cite{Richter}.   
Richter 
points out that, from
the point of view of physics,  the AP is ``an observation not an explanation".
Even if a certain physical property of our universe is a prerequisite for
life, the physicist must still show how it is implemented in our world and
how it is related to other properties and perhaps to an effective Lagrangian
of the world.  
Nonetheless, selection effects do constitute a form of at least partial
explanation for physical phenomena.
Presumably no experimentalist would deny that many deep
explanations in physics began with a seemingly coincidental observation. 
In particular, for the case at hand, it cannot be considered uninteresting that, 
billions of years before
life existed, the laws of physics were such as to allow the generation of life.
A sufficiently intelligent observer 
having only a sufficiently detailed understanding 
of the universe as it existed one second after the big bang would note that
there was, from at least that point on, a non-negligible probability for
the emergence of life.

   The string landscape \cite{landscape} postulates that there are a large number,
perhaps $10^{500}$, local minima of the effective potential each of which is
a potential universe with a set of particles, fundamental forces, and 
space time topology.  Because of the sheer number of these universes it is
proposed that there is a non-negligible probability that the universe would
have arrived at the one we observe in the requisite time and then life would
have arisen.  The string landscape proposal is, therefore, an attempt to
provide a physics explanation of the anthropic observation that the cosmological
constant and perhaps other properties of our universe are as they need to be
for the rise of life.  

     For the AP to be any guide and for the landscape to
be an efficient explanation, the number of life-supporting universes must be
reasonably small.  Thus, to understand our universe, it might be necessary to
seek alternatives that would still have allowed the evolution of life.   
Along these lines, it has recently been suggested \cite{Harnik}
that at least one alternative universe would have supported life, namely the
one with no weak interactions and other properties simultaneously altered to
allow nucleosynthesis and biochemistry.  If it is true that such a universe
would have allowed the evolution of life and if it is true that such alternatives
are in fact numerous then there would be no understanding as yet of why our
universe is as it is.  A single such alternative does not, of itself, render the
AP and the string landscape powerless since one does not necessarily
demand uniqueness but only a reasonable probability of landing in the observed 
universe.

     In our universe, as currently understood, the weak interactions play 
several important roles.  They probably provide the CP violation thought to be 
crucial in developing the baryon-antibaryon asymmetry essential to the
development of life.  The spontaneous breaking of the electroweak gauge group
allows for small Fermion masses.  An exact electroweak symmetry would have 
massless quarks and leptons.  A universe without weak interactions would be
expected to have elementary Fermion masses at the Planck scale. 
Weakly interacting neutrinos were important
in big bang nucleosynthesis and later in the stellar nucleosynthesis of heavy 
elements essential to life.  In our universe, they play a role
in stellar cooling and are crucial in the distribution of heavy elements
throughout the universe in Type II supernovae.  If one is to construct
theoretically a life-supporting weakless universe, all of these functions 
of the weak interactions in our universe need to be replaced by other 
mechanisms.  This is
precisely what the authors of ref.\,\cite{Harnik} have attempted to do.    
In addition, although not conclusively proven, it has long been proposed
that the parity violation of the weak interactions was important in
producing the essential left-right asymmetry of many organic molecules and 
of the human body, as reviewed for example in ref.\,\cite{Borchers}.
Although not the least likely explanation for the homochirality of life
\cite{Plaxco}, the weakness of the weak interactions pose challenges for
this hypothesis.  However,
it has also been proposed that the weak interaction effect on the 
homochirality of life was magnified in Type II supernovae \cite{Cline}.
If this connection becomes established a weakless universe would be 
definitively lifeless. 

     The proposed weakless universe has many other constants
that are tuned for life although these are "natural" in the sense of not
having quadratic renormalization once they are chosen as needed.  If we had
in fact become conscious in a weakless universe there would have been no
fewer mysteries than in our present one.  

However, beyond that, we question the assertion that the weakless universe
could in fact be life-generating.
As we remark in the next section, the weakless universe does not have an adequate
mechanism to distribute oxygen through the universe as would be needed to
make water and the four essential bases of DNA.  No one has succeeded at present
in presenting a plausible model of how life could exist without these.
Other potential problems with a weakless universe are also noted.

\section{\bf Oxygen deprivation in the weakless universe}
\setcounter{equation}{0}

    The authors of ref.\,\cite{Harnik} have chosen the primordial abundances in
such a way as to make deuterium plentiful in the early universe.  This allows them
a pathway to stellar nucleosynthesis of the heavy elements without the weak 
interactions that are so important in our universe.
In our universe, such heavy elements are dispersed through space
mainly by supernovae of two types: massive stars undergoing core collapse 
(Types II \& Ib) and  white dwarfs being pushed over their Chandrasekhar upper
mass limits by accretion in stellar binary systems (Type Ia).
In a weakless universe, as  ref.\,\cite{Harnik} acknowledge,
core-collapse supernovae will likely not occur,
because weak interactions are responsible for both the core collapse 
which initiates the explosion, as well as the energy deposition
which unbinds the outer layers of the star.
If core collapse supernovae require neutrino cooling and/or
neutrino energy deposition to proceed, there would be no core collapse
supernovae in a weakless universe.  
The absence of core collapse supernovae in a weakless universe
is crucial, because most of the oxygen in our universe comes from such supernovae.  

Therefore, in a weakless universe,
supernovae (Type Ia) from accreting white dwarfs  pushed over 
their Chandrasekhar mass limit would, as noted in ref.\,\cite{Harnik},
be the main source of heavy elements in the interstellar medium.
However, in the process of collapse, the carbon and oxygen in such
stars undergo rapid fusion to heavier elements and it is primarily
these heavy elements that emerge from the supernova.  In our universe
the elements ejected from Type Ia supernovae are dominated by iron,
in particular $^{56}$Fe, which results from the radioactive decay
of $^{56}$Ni (via $^{56}$Co) synthesized in the explosion.
A weakless universe lacks radioactivity, so such supernovae are
likely to generate more nickel than iron.

It is also unlikely that novae, thermonuclear flashes
from the surfaces of accreting white dwarfs, could oxygenate the universe.
The ejecta from novae are oxygen deficient (apart from producing
the relatively rare isotope $^{17}$O).  
They contain primarily 
the light elements, hydrogen
and helium, from the outer shells of white dwarfs. 
In addition, although novae are several thousand times more frequent than
supernova, they each eject only a millionth of the mass of a supernovae
\cite{Jose}.
Therefore,  we believe one can neglect novae as an adequate source of oxygen.

Even considering the full range of current theoretical numerical 
models for exploding white dwarfs \cite{Nomoto}-\cite{Iwamoto}, 
they produce only $\sim3-8\%$ of the oxygen mass, per supernova,
as does a typical core collapse supernova (as suitably averaged
over the yields and mass function of progenitors with a range
of masses).  Furthermore, to account for present day abundances
in our galaxy, historical core collapse supernovae have outnumbered 
SN Ia by a factor of $\sim5$ in galaxies like our own \cite{Iwamoto}.
Thus, even assuming neutrinoless alternative models are confirmed, 
less than about $1\%$ of our oxygen comes from SN Ia.

This oxygen deficiency in a weakless universe will strongly
inhibit the formation of planets and the development of life.
The incidence of extrsolar planets around nearby stars is
found to be strongly correlated with the iron abundance
of the host stars \cite{Fischer}: 25\% of sample stars with
an iron abundance of three times solar have planets, while
only 2\% of stars with one third solar iron abundance 
have planets.  Of 29 stars in the sample of ref.\cite{Fischer} 
with less than one third solar
iron abundance, none are observed to have planets.
Since the oxygen abundance is observed to be correlated 
with the iron abundance in such stars \cite{Ecuvillon}, 
and oxygen and iron are the dominant elements by mass in rocky
planets in the current universe, rocky planet
formation will be inhibited in a weakless universe.

Oxygen is thought to be particularly important for life
via liquid water, which provides a solvent to facilitate complex
chemistry.  The human body is $65\%$ oxygen by weight as are
the bodies of other mammals.  It is unlikely that a universe
without an abundant source of oxygen will evolve life.

The onset of star formation will be postponed
by two orders of magnitude of time in a weakless universe.
Since dark matter dominates the overall mass,
gravitational structure formation in the proposed
weakless universe will be largely unaltered.
But the density of baryons is 100 times less in
\cite{Harnik}, so the cooling time for the condensation
of baryons within dark matter halos will be 100 times
longer than in the current universe.
The radiative cooling time is
$$ t_c \approx {3kT \over n \Lambda_c}, $$
where $n^2 \Lambda_c(T)$ is the energy loss rate (per unit volume)
due to collisional cooling.  The temperature is set by
the gravitational potentials of dark matter halos, 
which are postulated to be the same in the proposed weakless universe,
but the gas densities $n$ are 100 times less than
in the current universe.  Thus, cooling times associated
with the onset of galaxy and star formation will be
extended by a similar factor, to $10^{10-11}$ years.
This delay is inconveniently close to the onset
of $\Lambda-$dominated acceleration of the universe.

The lack of radioactivity in a weakless universe provides
another challenge to the development of life.
Harnik, et al.\ acknowledged that the lack of radioactivity
would lead to very different evolution of rocky planets,
since radioactive heat drives vulcanism, but this was
not viewed as a serious obstacle to the model.
This may be a very serious obstacle, however, since
vulcanism is thought to be essential for maintaining
a stable greenhouse effect, provided
it is tempered by feedback processes due to the presence 
of liquid water \cite{Plaxco}.
Temperature stability on the Earth's surface is thought
to be due to atmospheric carbon dioxide levels set by 
an interplay between volcanic generation versus dissolution
and precipitation in liquid water oceans \cite{Walker}-\cite{Kasting}.
A weakless universe would have much less water 
and no radioactive heating, so the time scale for
rocky planets to have stable surface temperatures is
substantially reduced.  The interior heating of rocky 
planets will be only due to gravitational compression,
which will dissipate after a billion years.
In the present universe, the radioactive decay of
thorium and uranium provides heat on a timescale
10 times longer.

      In addition, to be considered on a par with the current universe
 and not introduce additional puzzles, the weakless
 universe should be part of a grand unified theory.  This is probably not
 possible unless the sum of the charges of the elementary consituents is zero
 unlike in the proposed weakless universe.   However, if this
 problem is overcome, perhaps
 by adding new charged particles at the GUT scale, then, after breaking,
 the strong
 and electromagnetic couplings should run according to
\be
  \frac{d\alpha^{-1}}{d \ln(M^2)} = - \frac{1}{3\pi} \sum Q_i^2
\ee
and
\be
   \frac{d\alpha_s^{-1}}{d \ln(M^2)} = \frac{33-2 n_f}{12\pi}
\ee
where the number of flavors is $n_f=6$ in our universe and $n_f=3$ in the proposed
universe.

Thus, at lowest order the couplings will satisfy
\be
    \alpha^{-1} &=& \alpha_0^{-1} - \frac{1}{3\pi} \sum Q_i^2 \ln({M^2/M_{X}^2})   \\
    \alpha_s^{-1} &=&\alpha_0^{-1} +\frac{33-2 n_f}{12\pi}  \ln({M^2/M_{X}^2}) \quad .
\ee

Here $\alpha_0$ is the unification coupling and $M_X$ is the unification mass.
These are strongly constrained by the requirement that the strong coupling constant
at low energies does not significantly differ between our and the weakless universe
in order to preserve the triple Carbon coincidence that is essential to produce
heavy elements.

If $\alpha_0^{-1}$ is much less than $\alpha^{-1}$ at low energies as holds in
our universe, then, independent of $M_X$ to lowest order,

\be
     \alpha_{\not w} = \frac{8}{3} \alpha \quad .
\ee

That is, the low energy fine structure constant in a weakless universe would be 
expected to be more than twice as strong as in our universe.  This would have
major effects on atomic and molecular energy levels and on the size of atoms
and thus cause significant
changes in chemical binding.  In addition, increasing the Coulomb barrier by
more than a factor of two would increase the temperature required for fusion.
To establish that the weakless universe would be hospitable to life would 
require much more analysis of these effects than has been undertaken in ref.\,\cite{Harnik}. 

\section{\bf Conclusions}
\setcounter{equation}{0}

     In ref.\,\cite{Harnik}, it is proposed that a universe without
weak interactions could support life.  If there are many hospitable
alternative universes, the anthropic principle based on observer
selection would no longer be a useful guide to understanding the
properties of our world.  In particular, the smallness of the 
cosmological constant, in both our universe and the weakless one,
would have no current explanation.  Faced with a plethora of
life supporting alternative universes, the string landscape ideas
\cite{landscape}
would also lose whatever predictive power they might have unless
the weakless universe or other alternatives could be shown not to
be among the local minima of string theory.

     Obviously, however, much further theoretical analysis would
be necessary to confirm that a universe without weak interactions
would, indeed, allow the evolution of life.  The nuclear reactions
proposed by the authors of ref.\,\cite{Harnik} as an alternate
mechanism for stellar nucleosynthesis would need to be studied in
greater detail.  The apparent fine tuning of quark 
abundances in a weakless big bang would have to be understood.

     In addition to such open questions, however, we have proposed
that a serious problem in a weakless universe from the point of view
of generating life is the difficulty of distributing oxygen through
the universe in anywhere near the required abundance.  Such a universe
would be extremely deficient in oxygen and life would,
in effect, have been suffocated at an early stage.  

     Our observations reduce the probability of generating life by at
least a factor of 100 below what is already thought to be a low number.  
If, however, the universe is infinite or many orders of
magnitude larger in size than the visible universe, it cannot be ruled out that
a statistical fluctuation in the oxygen abundance produces a
life sustaining region in the weakless universe.  In this case, a definitive proof
of the lifelessness of the weakless universe would have to await a proof
that the weak interactions are the sole feasible source of the homochirality
of life as discussed in the introduction.  Conversely a proof that a
weakless universe could support life would require a demonstration that
alternate sources of homochirality are effective and that the other
potential problems discussed in this note are avoidable.    

{\bf Acknowledgements}

    This work was supported in part by the US Department of Energy under
grant DE-FG02-96ER-40967.

\par
\end{document}